\documentclass[runningheads]{llncs}
\usepackage{algorithm}
\usepackage[noend]{algpseudocode}
\usepackage{xcolor}
\usepackage{color}
\usepackage{cite}
\usepackage{amsmath,amssymb,amsfonts}
\usepackage{array,graphicx}
\usepackage{listings}

\definecolor{textblue}{rgb}{.2,.2,.7}
\definecolor{textred}{rgb}{0.54,0,0}
\definecolor{textgreen}{rgb}{0,0.43,0}
\begin{document}

\title{HDR-Fuzz: Detecting Buffer Overruns using AddressSanitizer
  Instrumentation and  Fuzzing}
\author{Raveendra Kumar Medicherla \inst{1} \and 
 Malathy Nagalakshmi \inst{2} \and
Tanya Sharma \inst{2} \and
 Raghavan Komondoor\inst{3}}
\authorrunning{Raveendra et al.}
\institute{TCS Research, Tata Consultancy Services, Bangalore, India, 
\email{raveendra.kumar@tcs.com}
\and
Bangalore,India,\email{\{malathy124,tanya.sharma0217\}@gmail.com} 
\and
Indian Institute of Science, Bangalore, India, 
\email{raghavan@iisc.ac.in}
}
\maketitle
\begin{abstract}
Buffer-overruns are a prevalent vulnerability in software libraries and
applications.  Fuzz testing is one of the effective techniques to detect
vulnerabilities in general.  Greybox fuzzers such as AFL automatically
generate a sequence of test inputs for a given program using a
fitness-guided search process. A recently proposed approach in the
literature introduced a buffer-overrun specific fitness metric called
``headroom", which tracks how close each generated test input comes to
exposing the vulnerabilities. That approach showed good initial promise,
but is somewhat imprecise and expensive due to its reliance on conservative
points-to analysis. Inspired by the approach above, in this paper we
propose a new ground-up approach for detecting buffer-overrun
vulnerabilities.  This approach uses an extended version of ASAN (Address
Sanitizer) that runs in parallel with the fuzzer, and reports back to the
fuzzer test inputs that happen to come closer to exposing buffer-overrun
vulnerabilities.  The ASAN-style instrumentation is precise as it has no
dependence on points-to analysis. We describe in this paper our approach,
as well as an implementation and evaluation of the approach.

\end{abstract}
\section{Introduction}
\label{sec:introduction}
\emph{Software vulnerabilities} are  flaws in software that can be 
exploited by an attacker to gain control over the system.  
Buffer overruns are a very prevalent ``implementation vulnerability'' in
real world software. Buffer overrun is the third most prevalent type of
vulnerability as per the CVE database~\cite{BOFLOW2017}. A buffer-overrun
occurs when a program erroneously writes to a buffer (or array) beyond the
limits of the allocated buffer (or array). An attacker can exploit this vulnerability to take control of the program, for example by overwriting the return address of a function on the stack.  

Detecting vulnerabilities such as buffer overruns is a key requirement for
building a secure software~\cite{SurveyVulnerability}.  \emph{Fuzz
  testing}~\cite{Miller_1990} is an automated technique that aims to
uncover vulnerabilities by generating new test inputs by \emph{fuzzing}
(i.e., mutating) already generated test inputs, and by looking for bug-inducing
inputs among this generated sequence of test inputs. \emph{Greybox fuzzing}
is a specific kind of fuzz testing that is employed in several practical
tools and approaches~\cite{Rawat2017,AFL,bohme2017directed}.  It searches
for bug-inducing inputs using a fitness-guided search process, and using
lightweight instrumentation in the program  to compute the fitness
of each generated test input from the (instrumented) run on this input. 

Most existing greybox fuzzers, such as the industry-standard tool
AFL~\cite{AFL}, retain a newly generated test-input if a run using it
\emph{covers} a new region of code. However, better coverage alone need not
result in better detection of specific kinds of vulnerabilities.
Recognizing this, an approach and tool called AFL-HR~\cite{rr2020sbst} was
described recently, which introduced the novel notion of \emph{headroom} as
a fitness metric for buffer overruns (and other related
vulnerabilities). This metric assigns higher fitness (and hence causes
the retention of) test inputs that come \emph{closer} to causing a buffer
overrun than other test inputs generated so far at any buffer write
location in the program. The approach mentioned above was implemented as an
extension over  AFL, and showed significant
improvement over baseline AFL in its ability to identify real buffer
overrun vulnerabilities. However, this approach has two key
limitations. The first is its reliance on static points-to analysis to
instrument buffer write locations, which can be impractical to use on
large applications, and imprecise as well.  The second limitation is that
the headroom instrumentation is expensive, and thus the rate at which AFL
executes the (headroom) instrumented program slows down. In this paper we
propose an entirely new greybox fuzzing approach, which we call
\emph{HDR-Fuzz}, which still uses the headroom  metric, but
 eliminates both these limitations.

The novelty of HDR-Fuzz is in its architecture. The architecture consists
of two parallel processes.  The first process is a modified version of AFL
that uses only AFL's original, lightweight coverage instrumentation, and
relies on the second process to compute the headroom of any test input.
The first process passes a selected subset of test inputs that it generates
to the second process via a queue. The second process is based on a custom
extension of Address Sanitizer (ASAN)~\cite{ASAN2012} that we have
devised. ASAN is an industry-standard tool to \emph{detect} if a
buffer-overrun occurs in a given run. We have extended ASAN's
instrumentation regime to calculate the headroom of the run at each buffer
write location that was visited but not overflowed in the run. This
ASAN-based process needs no static pointer analysis, and is hence
applicable practically on a large set of real programs. The second process
calculates the headroom at each buffer write location due to each test
input it receives, and communicates back test inputs that achieve good
headroom-based fitness to the first process. The first process then retains
these test inputs (in addition to test inputs that achieve better coverage)
for further fuzzing.

We describe in this paper our approach as well a prototype tool based on
the approach. We also describe initial experimental results from our tool on
three standard suites of benchmarks. The results indicate substantial performance gains of our tool against the
baseline AFL as well as the recent tool AFL-HR~\cite{rr2020sbst}.

\section{Background}
In this section, we provide brief background on the existing tools ASAN and
AFL that is key to understanding our approach.
\subsection{AddressSanitizer}
\label{ssec:backg:asan}
AddressSanitizer (ASAN)~\cite{ASAN2012} is an instrumentation based
memory-error detector that detects various types of memory-related
errors. We restrict our attention to its buffer-overrun detection
mechanism. The ASAN instrumentation appends fixed-size extra regions of
memory, called \emph{red zones}, around both sides of every buffer in the
\emph{main} (i.e., actual) memory (whether the buffer is a global, or is on
stack or in heap).  ASAN also maintains a separate \emph{shadow memory},
wherein there is a single byte corresponding to every eight bytes in the
main memory.   Shadow memory
locations corresponding to actual in-use memory locations contain value 0,
whereas shadow memory locations corresponding to red zones in the main
memory contain negative values.  When any memory lookup happens, ASAN uses
an efficient address translation scheme to translate the main-memory address to
its corresponding shadow-memory address, and reports an overrun if there is
a negative value at the shadow memory location.

\subsection{AFL}
\begin{algorithm}[!ht]
\caption{Enhanced AFL Test generation}
\label{approach:afl}
\small
\begin{algorithmic}[1]
\Require Given program $P_a$ with AFL's coverage instrum., and a ``seed'' input $s$.
\Ensure  A tree of test inputs $T_G$ for $P_a$
\State   Create an empty tree $T_G$ of test inputs
\State   Initialize queues $Q_A$, $Q_R$ to empty\label{alg:hdr:qa}
\State   $t_r.\mathit{data}$ = $s$\Comment{$t_r$ is a new tree node}
\State   $T_G.setRoot(t_r)$
\Repeat\label{alg:outerloop}
	\State $\langle t_h,S_h \rangle$ = \textproc{dequeue}($Q_R$)\label{alg:hdr:beg1}
	  \ForAll {$t_a$ \textbf{in} $S_h$} 
		\If{$t_a \not\in T_G$}
			\State \textproc{addChild}($T_G$,$t_{h}$,$t_a$) \label{alg:hdr:add1}		
        \EndIf
	 \EndFor \label{alg:hdr:end1}
 \State Let $t$ = \textproc{selectNextH}($T_G$)\label{alg:hdr:select}
 \State Let $N$ = \textproc{getFuzzPotentialH}($t$)\label{alg:hdr:score}
 \State Let $T_n$ = \textproc{generateOffspring}($t$,$N$) \label{alg:hdr:gen}
 \ForAll {$t_g$ \textbf{in} $T_n$}
    \State Let $I_g$ = \textproc{run}($P_a$, $t_g.\mathit{data}$)\label{alg:hdr:exec}
		\If{\textproc{isFit}($I_g$)}\label{alg:hdr:isfit} %
       \State \textproc{addChild}($T_G$,$t$,$t_g$) \label{alg:hdr:add2}
   \EndIf
 \EndFor 
 \State Let $S_n$ = \textproc{SampleInputs}($T_n$) \label{alg:hdr:samp}
 \State \textproc{enqueue}($Q_A$,$\langle t,S_n \rangle$) \label{alg:hdr:enq2}	
\Until{\emph{user terminates the run}}
\State \Return $T_G$
\end{algorithmic}
\end{algorithm}
 AFL is a fitness-guided greybox fuzzer geared to discover
test inputs that maximize code coverage. Algorithm~\ref{approach:afl} gives
a high-level and simplified overview of AFL. For the current discussion,
ignore
Lines~\ref{alg:hdr:qa},~\ref{alg:hdr:beg1}--\ref{alg:hdr:end1},~\ref{alg:hdr:samp},
and~\ref{alg:hdr:enq2}, as they are not part of the original AFL.  Given an
instrumented program $P_a$ that carries AFL's coverage-based
instrumentation, 
and an initial test input $s$, the approach iteratively generates test
inputs, and \emph{retains} some of them in a tree $T_G$. Each node of the
tree corresponds to a test input, and its ``\emph{data}'' field contains
the actual test input (a sequence of bytes). The algorithm then works
iteratively and continually using the loop that begins at
Line~\ref{alg:outerloop}, by selecting a test input $t$ from $T_G$ for
fuzzing in each iteration (Line~\ref{alg:hdr:select}). 

The \emph{fuzzing potential} of a test input $t$ is a number $N$, computed
heuristically in~Line~\ref{alg:hdr:score} by invoking a
sub-routine~\textproc{getFuzzPotentialH}. In Line~\ref{alg:hdr:gen}, AFL
generates $N$ test inputs as offspring of $t$ (by fuzzing $t$.\emph{data}).
AFL uses different \emph{genetic operators} such as \emph{flipping bits} at
specific locations in $t$.\emph{data}, \emph{copying bytes} from one
location in $t$.\emph{data} and writing them to some other location, etc.

Each test input generated in Line~\ref{alg:hdr:gen} is executed using AFL's
coverage-based instrumentation in Line~\ref{alg:hdr:exec}. The result
of the run is a \emph{coverage profile}, which is stored in the structure
$I_g$. 
Line~\ref{alg:hdr:add2} of Algorithm~\ref{approach:afl} adds the newly
generated test input $t_g$ to $T_G$ as a child of $t$ if $t_g$ has
\emph{significantly different} coverage profile than the profiles of test
inputs that are already in $T_G$. This ``significant difference'' check is
done by routine~\textproc{isFit}, which is called in
Line~\ref{alg:hdr:isfit}. Due to space limitations we are compelled to
omit many details from the discussion above.

\section{Our approach}
\label{sec:appr}
In this section, we describe our approach in terms of our changes to AFL, 
our enhancements to ASAN to yield headroom-checking instrumentation,  and
finally, the \emph{Driver} component of our tool, which uses headroom
instrumentation to identify test inputs that have good fitness based on
the headroom metric. 
Note that (our modified version of) AFL and the Driver are the two parallel
processes that constitute our overall system.

\subsection{AFL enhancements}
\label{ssec:AFL}

Lines~\ref{alg:hdr:qa},~\ref{alg:hdr:beg1}--\ref{alg:hdr:end1},~\ref{alg:hdr:samp},
and~\ref{alg:hdr:enq2} in Algorithm~\ref{approach:afl} describe our
enhancements to AFL to make it utilize headroom checking.
Lines~\ref{alg:hdr:samp}-\ref{alg:hdr:enq2} sample a subset $S_n$ of the
set of newly generated test inputs $T_n$, and send them (along with their
parent test input $t$) to the driver via a shared queue $Q_A$. Sampling is
necessary, as running every generated test input with headroom
instrumentation within the Driver would be prohibitively expensive. Our
sampling selects all test inputs that are already in $T_G$, and
additionally a random subset of test inputs from $T_n - T_G$ (i.e., test
inputs that AFL chooses not to retain due to insufficient new
coverage). The size of this subset (as a percentage) is a parameter to the
approach. The intuition behind selecting inputs in $T_n - T_G$ is that
though a test input may be uninteresting from a coverage perspective, it
may come closer to exposing a vulnerability.

We will discuss the working of the driver in more detail in
Section~\ref{ssec:driver}. For each record $\langle t, S_n \rangle$ that
the driver receives from the queue $Q_A$, it returns to
Algorithm~\ref{approach:afl} the subset of $S_n$ consisting of test inputs
that are fit wrt the headroom metric, along with the parent test input
$t$. Algorithm~\ref{approach:afl} adds the test inputs received from the
driver to $T_G$ (see Lines~\ref{alg:hdr:beg1}--\ref{alg:hdr:end1} in
Algorithm~\ref{approach:afl}).  The dequeing operation in
Line~\ref{alg:hdr:beg1} is non-blocking, and returns an empty set $S_h$ in
case the queue is empty.

\textproc{selectNextH} (called in Line~\ref{alg:hdr:select}) works as
follows.  As it is invoked repeatedly throughout the run of the algorithm,
it alternatively selects test inputs that were added to $T_G$ in
Lines~\ref{alg:hdr:add1} and~\ref{alg:hdr:add2} of the algorithm. 
The intuition is that we would like to give equal priority to
fuzzing test inputs that attain new coverage (in order to cover all parts
of the program well) and to fuzzing test inputs that reduce headroom (in
order to come closer to exposing vulnerabilities in parts of the program
that are already covered).

After Algorithm~\ref{approach:afl} is terminated, the vulnerability
exposing test inputs are reported by picking up test inputs in $T_G$ that
cause crashes when the  program is run using (normal) ASAN instrumentation. 

\subsection{Headroom-checking instrumentation}
\label{ssub:ASAN}
Our headroom-checking instrumentation piggy-backs on ASAN's
instrumentation. The \emph{raw headroom}~\cite{rr2020sbst} at a
buffer write location due to a run  measures how close (in
terms of number of bytes) the buffer write pointer came to the end of the
buffer across all visits to the location in the run. The \emph{scaled headroom} 
is the ratio of raw headroom over the size of the buffer multiplied by 128. A low value of
headroom means that the run came close to exposing a buffer overrun. 

\begin{figure}
  \begin{scriptsize}
	\begin{minipage}{\textwidth}
   \begin{tabular}{ll}
\begin{minipage}{.5\textwidth}
\begin{verbatim}
// Instrumented program has a global 
// array called headroom_array, 
//indexed by buffer-write locations.

calculate_headroom(uptr addr, int storeIdx) {
  // storeIdx is location of currently visited
  // buffer-write instruction.
  // addr is memory address used in this visit.
  writeAddr = getShadowAddress(addr);
  endAddr = begAddr = writeAddr;
  int rawHeadroom = leftMargin = 0;
  while (endAddr is not in right redzone) { 
    // Move right 
\end{verbatim}
\end{minipage}
&
\begin{minipage}{.5\textwidth}
\begin{verbatim}
    endAddr = endAddr + 1;	
    rawHeadroom = rawHeadroom + 1;
  }
  while (begAddr is not in left redzone) {
     // Move left 
     begAddr = begAddr - 1;	
     leftMargin = leftMargin + 1;
  } 
  byte scaledHeadroom = (rawHeadroom * 128) / 
               (leftMargin + rawHeadroom - 1);
  headroom_array[storeIdx] = scaledHeadroom;
}
\end{verbatim}
\end{minipage}	
\\		
\end{tabular}

	\end{minipage}
	\end{scriptsize}
   \caption{Headroom calculation at a buffer-write event}
   \label{fig:psuedo1}
\end{figure}

Our instrumentation calculates the scaled headroom due to the run at every
\emph{buffer write} location, whether the write location is a global,
stack, or heap buffer. 
Figure~\ref{fig:psuedo1} depicts the pseudo code of routine
\emph{calculate\_headroom}, which our instrumentation invokes from each
buffer-write location. The parameter \emph{addr} is the address at which
the current write is happening, while \emph{storeIdx} is the Location ID
(or program counter) of the current write instruction. The
subroutine~\emph{getShadowAddress} is ASAN's address translation routine
(see Section~\ref{ssec:backg:asan}). \texttt{headroom\_array} is a global array
indexed by buffer-write locations, which, at the end of each run of the
program, stores the scaled headrooms  at the buffer write
locations that were visited in the run. 
%(all entries of this array are initialized to 128 at
%the start of the run). 
We refer to the contents of this array at the end of
a run as the \emph{headroom profile} of the run. The rest of the pseudo-code in
Figure~\ref{fig:psuedo1} is self-explanatory.

\subsection{Driver}
\label{ssec:driver}
\begin{algorithm}[!ht]
\small
\caption{Coordinating driver}
\label{approach:driver}
\begin{algorithmic}[1]
\Require Given program $P_h$ with headroom instrumentation \Ensure Set of
test inputs $S_h$ to be retained due to headroom.  \State Initialize array
\emph{minHProfile} with value 128 in
all entries.
\Repeat\label{alg:drvr:outerloop1}
	\State $\langle t,S_n \rangle$ = \textproc{dequeue}($Q_A$) \label{alg:drvr:deque}
	  \State $S_h  = \emptyset $ 
		\ForAll {$t_s$ \textbf{in} $S_n$}
    \State Let $I_h$ = \textproc{run}($P_h$, $t_s.\mathit{data}$)\label{alg:drvr:exec}
			\If{\textproc{isLess}($I_h$, \emph{minHProfile})}\label{alg:drvr:isfit}
					\State Add $t_s$ to $S_h$  \label{alg:drvr:add}
                    \State \emph{minHProfile} =
                    \textsc{min}(\emph{minHProfile}, $I_h$)
			\EndIf			
		\EndFor
		\If{$S_h \ne \emptyset$}
		  \State  \textproc{enqueue}($Q_R,\langle t,S_h \rangle$) \label{alg:drvr:enq}
		\EndIf	
\Until{\emph{user terminates}}
\State \Return
\end{algorithmic}
\end{algorithm}

Algorithm~\ref{approach:driver} describes the driver, which runs in
parallel with Algorithm~\ref{approach:afl}.  The given program $P_h$
carries our headroom instrumentation as described in
Section~\ref{ssub:ASAN}. The algorithm maintains a globally minimum
headroom profile across all runs of the program in the global array
\emph{minHProfile}, indexed by the buffer-writing locations in the program.
The algorithm waits for test inputs from Algorithm~\ref{approach:afl} in
the queue $Q_A$ (Line~\ref{alg:drvr:deque}). On each received test input
$t_s$, it runs the instrumented program $P_h$ in Line~\ref{alg:drvr:exec},
and picks up the headroom profile at the end of the run and puts it into
the temporary array $I_h$. The function \textproc{isLess} in
Line~\ref{alg:drvr:isfit} checks whether the headroom in any entry of $I_h$
is less than the corresponding entry of \emph{minHProfile}; if yes, $t_s$
is saved into a set $S_h$. In Line~\ref{alg:drvr:enq}, all the saved test
inputs are sent back to AFL via the queue $Q_R$.

\section{Implementation and Evaluation}
We have implemented our approach as a tool named
HDR-Fuzz. 
%It is currently available for download here:
%{\scriptsize https://drive.google.com/file/d/1jOtU4x02lXaFBYMxpYNvKRTCCxFURPlw}. 
Our implementation is built on top of AFL 2.52 and ASAN 10, and uses the C
and C++ languages.

\sloppypar We evaluate our tool on three buffer-overrun benchmark suites. The first
suite is a set of eight programs from the ``MIT Benchmarks'', which have
been used by previous researchers to evaluate buffer overflow detection
tools~\cite{Zitser2004}.
%, which have been used by previous researchers to
%evaluate a static-analysis and symbolic-execution based technique for
%detecting buffer overruns.
We use the following MIT benchmark programs: s1, s3, s4, s5, b1, b3, b4,
and f1.  The second suite is a set of 10 programs from the ``Cyber Grand
Challenge'' (CGC) benchmarks~\cite{Stephens2016}, which have been used to
evaluate fuzzers and symbolic execution tools. We use the following 10 CGC
benchmarks: {\small CROMU\_00030, CROMU\_00084, CROMU\_00088, KPRCA\_00060,
  TNETS\_00002, KPRCA-00001, CROMU\_00041, CROMU\_00020, KPRCA\_00041}, and
{\small KPRCA\_00045}.  The third suite contains nine programs from
Google's ``fuzzer-test-suite''~(FTS)~\cite{GFTS}.  These benchmarks are
much larger in size, and are derived from real applications or libraries.  We
use the following nine FTS benchmarks: {\small openssl-1.0.1f,
  libxml2-v2.9.2, libarchive-2017-01-04, woff2-2016-05-06,
  openthread-2018-02-27-rev1, openthread-2018-02-27-rev2,
  openthread-2018-02-27-rev7, openthread-2018-02-27-rev8/9/10}, and {\small
  openthread-2018-02-27-rev11}.  We compared the performance of our tool
with (standard) AFL. We used Google cloud ``standard'' machines, with 8 hardware
threads and 32GB memory, for our evaluations. For each benchmark program,
we ran both AFL and our tool HDR-Fuzz on the benchmark
 (starting from a small manually created ``seed'' test input). To
mitigate the effects of randomness in the fuzzers, we ran both tools on
each benchmark program three times, with a 3-hour time budget for each
run.

\subsection{Experimental results on MIT and CGC benchmarks}
We present our experimental results on the MIT and CGC benchmarks together,
as both these suites consist of small to medium size programs.  The MIT and CGC benchmarks
together have 61 known vulnerable buffer overrun locations (49 in MIT, and
12 in CGC). These locations are indicated by the suite designers, but this
information is (obviously) not given by us to either tool.

%% between them. Each of these locations are marked in the source program by the suite designers.  Our results are summarized in Table~\ref{tab:mit}. In this table, Column~2 shows the benchmark program name and Column~3 shows the number of vulnerabilities that are marked by the suite designers. Column~4 shows the number of locations instrumented by ASAN. Column~5 shows the number of vulnerabilities that are detected by AFL and column~6 shows the vulnerabilities detected by our tool.

%% \begin{table}[h]
%% \caption{Experimental Results with MIT Benchmark}
%%   \centering
%% \small  
%% 	\setlength\tabcolsep{3pt}
%% \begin{tabular}{|c|c|c|c|c|c|}
%%   \hline
%%   % after \\: \hline or \cline{col1-col2} \cline{col3-col4} ...
%%   S.No & Benchmark & Potential  & \# Instru- & \multicolumn{2}{c|}{No. of Detected}  \\
%%        &           & overflow & mented    & \multicolumn{2}{c|}{Overflows}   \\ \cline{5-6}
%%        &           & locations  & locations & AFL & HDR-Fuzz \\
%% 	\hline		
%%   1    & s1 & 28 & 29 & 9 & 24 \\
%%   2    & s3 & 3  & 14  & 1  & 3 \\
%%   3    & s4 & 7  & 13  & 4  & 5 \\
%%   4    & s5 & 3  & 15  & 1  & 1 \\
%%   5    & b1 & 1  & 35  & 1  & 1 \\
%%   6    & b3 & 1  & 4   & 0  & 1 \\
%%   7    & b4 & 2  & 18  & 1  & 1 \\
%%   8    & f1 & 4  & 4   & 1  & 2 \\
%%   \hline
%% \end{tabular}
%% \label{tab:mit}
%% \end{table}
%\begin{center}Fig. 4. Results with MIT Benchmark\end{center}
%\begin{center}Table 1: Average number of buffer overflow violations detected by each tool with MIT Benchmark\end{center}
%\includegraphics[width=\columnwidth]{MIT-table-new.png}

\begin{figure}[t]
\includegraphics[width=0.75\textwidth]{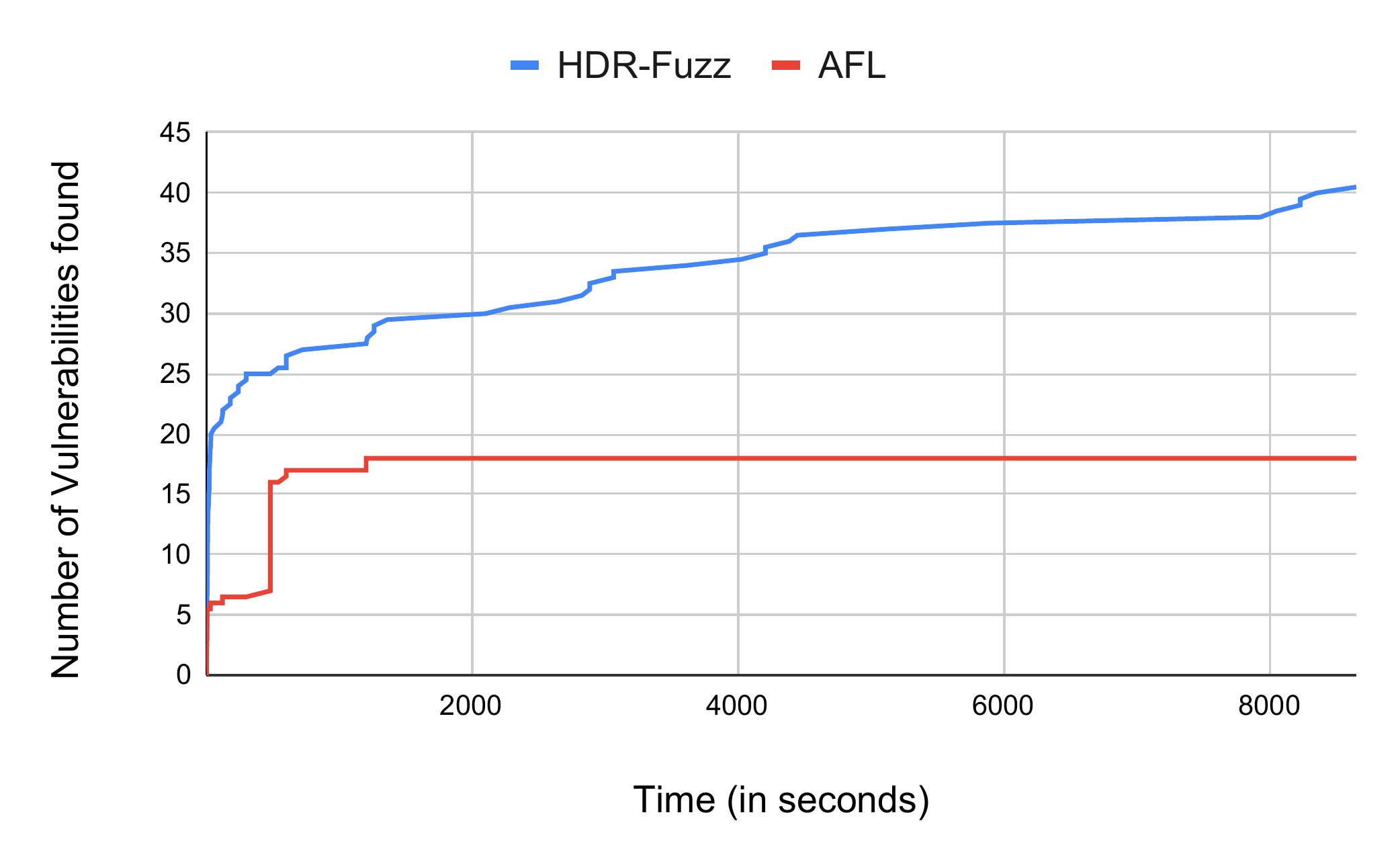}
\caption{Results from MIT and CGC Benchmarks}
\label{fig:mit}
\end{figure}

Figure~\ref{fig:mit} shows the cumulative number of vulnerabilities found
over time across all the eighteen benchmarks by AFL and by our tool. At any
point of time $t$, for each benchmark, we have considered the average
number of vulnerabilities exposed up to that point of time across the three
runs, added up the averages corresponding to all eighteen benchmarks, and
plotted this sum against point $t$ in Figure~\ref{fig:mit}.  Overall, our
tool detects an average of 40.5 (out of 61) vulnerabilities, which is more
than twice compared to AFL, which detects only 18. From the graph, it is
also clear that our tool finds vulnerabilities much more quickly; e.g.,
within the first 200 seconds our tool has found more than 20
vulnerabilities while in the same time AFL has found only about 7
vulnerabilities. Note, we cutoff the plots at 9000 seconds (without going
all the way to 3 hours) because no new vulnerabilities were found by either
tool beyond the cut-off point chosen.

\subsection{Experimental results on FTS benchmarks}
\begin{figure}[t]
\includegraphics[width=0.75\columnwidth]{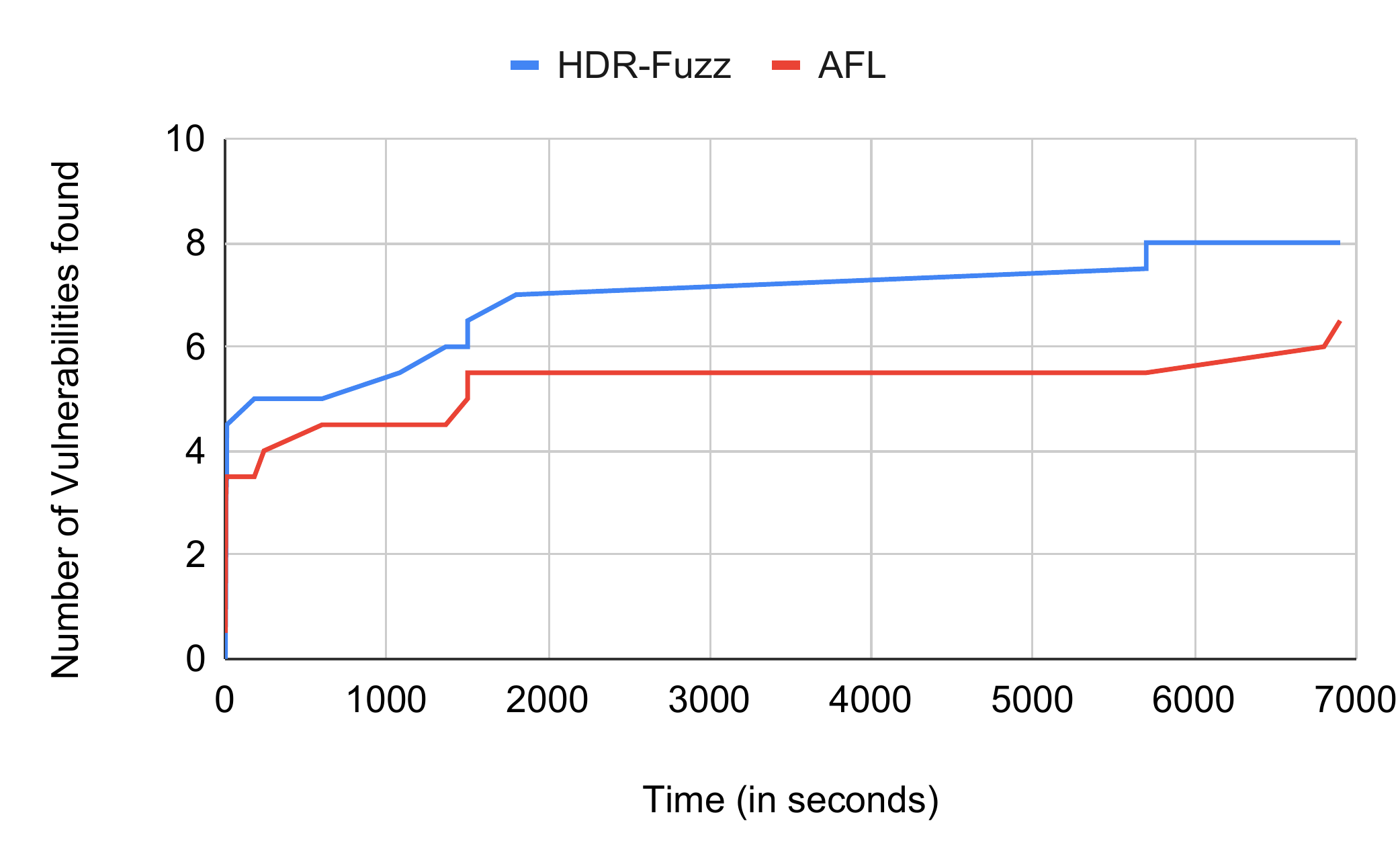}
\caption{Results from FTS benchmarks}
\label{fig:fst}
\end{figure}
%\begin{center}Table 2: Average number of buffer overflow violations detected by each tool with Fuzzer Test Suite Benchmark\end{center}
%\includegraphics[width=\columnwidth]{fuzzer_table.png}

We followed the same experimental procedure with the FTS benchmarks as with
the MIT and CGC benchmarks.  Figure~\ref{fig:fst} depicts the results (in the same
manner as Figure~\ref{fig:mit}). There are a total of 11 known
vulnerability locations across the nine benchmarks. In summary, our
approach detects an average of 8 vulnerabilities, while AFL
detects an average of 6.5 vulnerabilities. That is, our tool finds 23\%
more vulnerabilities. Our tool is quicker here as
well, as the gap in vulnerabilities found by the two tools arises very
early in the runs itself. Although the gap between the two tools is not as
large as with the MIT benchmarks, it is still quite significant.

\subsection{Comparison with AFL-HR}
\label{ssec:comparison-with-afl}
We briefly compare here the performance of our approach with the recent
approach AFL-HR, which originally introduced the notion of headroom. Our
comparison is based on the results as reported in their
paper~\cite{rr2020sbst}. They had applied their approach only on the MIT
benchmarks. Their tool had detected 29 vulnerabilities on average (out of
49 total) within the same 3-hour time budget, whereas our  tool
detects approximately 38.5 vulnerabilities on average on these same
benchmarks. This is a substantial improvement in performance. We believe
the improvement is due to two reasons: Firstly, AFL executes more runs per
unit time in our approach due to the absence of headroom instrumentation in
the version of the program given to AFL. Secondly, our driver process
(based on ASAN) calculates headroom directly and precisely at all buffer
write locations using memory-instrumentation. Whereas AFL-HR's headroom
computation instrumentation is based on (imprecise) static points-to
analysis, and calculates headroom precisely only at buffer write locations
where the buffer write pointer is guaranteed to point to a unique
source-level declared buffer across all visits to the location in all runs.

\section{Related work}
\label{sec:related}
Due to space limitations, we compare our work only with very closely
related work. We have taken the idea of the headroom metric from the recent
approach AFL-HR~\cite{rr2020sbst}. However, our tool architecture is quite
different from AFL-HR's, and is based on two communicating and
collaborating processes as opposed to a single process.  Also, our 
tool has no dependence on static points-to analysis. The improvement in
effectiveness of our approach due to these factors was discussed in
Section~\ref{ssec:comparison-with-afl}. Also, due to the independence from
static points-to analysis, our approach is applicable on large benchmarks
like FTS, on which it is not clear whether AFL-HR would scale.

In recent years there has been a large body of reported work on greybox
fuzz testing, as this approach has been found to be scalable and
practical. Basic coverage-based greybox fuzzing approaches were proposed
first~\cite{AFL,fraser2012whole}. Subsequently, researchers have proposed
extensions such as to prioritize the coverage of low-frequency
paths~\cite{Stephens2016,Boehme2016,Rawat2017,chowdhury2019verifuzz}, and
to direct fuzzers to reach more quickly a given set of target program
locations\cite{bohme2017directed}. These approaches are not adept at
exposing complex vulnerabilities such as buffer overruns that get exhibited
only in runs that reach vulnerability locations with certain specific
vulnerability-inducing program states.  FuzzFactory~\cite{fuzzfactory2019}
is a framework for instantiating a fuzzer with domain-specific testing
objectives. However, their approach does not focus on detecting
vulnerabilities, and does not have the notion of how close a test run comes
to exposing a vulnerability.

\section{Conclusions and Future Work}
In this paper, we proposed an approach that combines
AFL's fuzzing and headroom instrumentation based on ASAN's shadow memory to
detect buffer overrun vulnerabilities effectively and efficiently.  Our tool
was robust and applicable to very large benchmarks, and detected many
more vulnerabilities than baseline AFL as well as the recent tool AFL-HR. 

In future work, we would like to extend our approach to identify other
kinds of vulnerabilities that ASAN currently targets such as use-after-free
errors, as well as other types of vulnerabilities such as integer overflows
and assertion violations for which we might need to use runtime tools
beyond ASAN.

%\bibliographystyle{splncs04}
%\bibliography{fuzzing}
\end{document}